# Structurally complex Frank–Kasper phases and quasicrystal approximants: electronic origin of stability

Valentina F Degtyareva* and Natalia S Afonikova

Institute of Solid State Physics, Russian Academy of Sciences, Chernogolovka 142432, Russia
* Correspondence: degtyar@issp.ac.ru

**Abstract:** Metal crystals with tetrahedral packing are known as Frank–Kasper phases with large unit cells with the number of atoms from hundreds to thousands. The main factors of the formation and stability of these phases are the atomic size ratio and the number of valence electrons per atom. The significance of the electronic energy contribution is analyzed within the Fermi sphere – Brillouin zone interactions model for several typical examples: $Cu_4Cd_3$, $Mg_2Al_3$ with over thousand atoms per cell, and for icosahedral quasicrystal approximants with 146 to 168 atoms per cell. Our analysis shows that to minimize the crystal energy, it is important that the Fermi sphere (FS) is in contact with the Brillouin zones that are related to the strong diffraction peaks: the zones either inscribe the FS or are circumscribed by the FS creating contact at edges or vertices.

## 1. Introduction

Common metallic structures are based on high-symmetry atomic cells such as face-centered cubic (*fcc*), close packed hexagonal (*hcp*) and body-centered cubic (*bcc*) that have a coordination number either 12 or 8+6. These structures are found in elements in their metallic phases; they are also formed in binary alloys and compounds if constituent elements have small differences in the atomic size and electronegativity. A special family of intermetallic alloys is based on tetrahedrally-packed phases, called Frank-Kasper phases, where tetrahedra form polyhedra with coordination number 12 (icosahedron), 14, 15, and 16 [1]. Frank-Kasper polyhedra are basic structural units for many types of metallic alloys including very complex structures with the number of atoms in the unit cell up to hundred or thousand, as well as quasicrystals. Structures and properties of complex metallic alloys are considered in several books and review papers [2-7].

Crystal structures of very complex intermetallics are built with successive shells of atoms including polyhedra like icosahedron, dodecahedron, rhombic triacontahedron, and truncated icosahedron (soccer ball or "fullercage" [8]). These atomic arrangements are called "Russian doll" ("matryoshka") clusters [9]. Similar atomic packings exist in icosahedral quasicrystals known as Mackay-type, Bergman-type and Tsai-type. These clusters build the structures of quasicrystal approximants with the periodic arrangements in the *bcc* cell (1/1 type) or more complex periodic structures.

Quasicrystal and approximant phases exist at certain alloy compositions usually defined by valence electron concentrations and represent a kind of Hume-Rothery phases [2-5]. The crystal energy is lowered by contact of the Fermi level to Brillouin zone planes formed by strong diffraction peaks. Historically Hume-Rothery phases were considered first in the Cu-Zn alloys and related binary systems of Cu-group elements with the neighboring elements of higher valences [10-12]. Classical Hume-Rothery phases with fcc → bcc → complex -brass → hcp are defined by the number of valence electrons per atom such as 1.35 → 1.5 → 1.62 → 1.75. In the case of quasicrystal approximants these values are usually higher. Diffraction patterns of these compounds consist of several strong diffraction peaks that should be taken into account when considering the stability of their crystal structures.

In the present paper the model of the Fermi-sphere – Brillouin zone (FS-BZ) interaction is applied to complex structures considering of the FS inscribed into BZ, as well as the FS enveloped the inner zones contacting to edges or vertices. Both cases of FS-BZ configuration should affect the band-structure energy and decrease the crystal energy.

## 2. Theoretical background and method of analysis

Formation of binary compounds at a certain alloy composition is defined by some important factors characterizing the alloy constituents, such as the difference in atomic sizes, electronegativity etc. Beyond these factors, formation of metallic structures is defined by effects of the Fermi sphere – Brillouin zone (FS-BZ) interaction. The Hume-Rothery mechanism has been identified to play a role in the stability of structurally complex alloy phases, quasicrystals and their approximants [2,13-16]. Formation of the complex structures of elemental metals under pressure can also be related to the Hume-Rothery mechanism [17-19]. Phase diagrams of binary alloys such as Au-Cd represent several phases with complex structures that follow the Hume-Rothery rule of stability at certain electron concentrations [20].

The band structure contribution to the crystal structure energy can be estimated by analyzing configurations of Brillouin-Jones zone planes within the nearly free-electron model. A special program BRIZ has been developed [21] to construct FS-BZ configurations and to estimate some parameters such as the Fermi sphere radius ($k_F$), values of reciprocal wave vectors of BZ planes ($q_{hkl}$) and volumes of BZ and FS. The BZ planes are selected to match the condition $q_{hkl} \approx 2k_F$ that have a significant structure factor. In this case an energy gap is opened on the BZ plane leading to the lowering of the electron band energy. The ratio of ½ $q_{hkl}$ to $k_F$ is usually less than 1 and equals ~0.95; it is called a "truncation" factor. In the FS-BZ presentations by the BRIZ program the BZ planes cross the FS, whereas in the real system the Fermi sphere is deformed and accommodated inside BZ due to an increase of the electron density and to a decrease in the electron energy near the BZ plane.

The crystal structure of a phase chosen for the analysis by the BRIZ program is characterized by the lattice parameters and the number of atoms in the unit cell, which define the average atomic volume ($V_{at}$). The valence electron concentration (z) is the average number of valence electrons per atom that gives the value of the Fermi sphere radius $k_F = (\ z/V_{at})^{1/3}$. Further structure characterization parameters are the number of BZ planes that are in contact with the FS, the degree of "truncation" factor and the value of BZ filling by electronic states, defined as a ratio of the volumes of FS and BZ. Presentations of the FS-BZ configurations are given with the orthogonal axes with the following directions in the common view: a* is looking forward, b* to the right and c* upward.

The FS-BZ interaction for relatively simple structures such as Cu-Zn phases is usually described as a contact of the FS to the BZ planes resulting in the decrease of the electronic energy by the formation of the energy gap. In the case of complex structures, diffraction patterns consist of several strong reflections at wave vectors well below $2k_F$ forming BZs that are inscribed by the FS with contacts at the edges or vertices. These cases are constructed with the BRIZ program by proper choice of presentations: "Sphere inside Polyhedron" or "Polyhedron inside Sphere". Structural stabilization due to the electronic band contribution where FS is enveloping the BZ was discussed for phase stability in In and Sn alloys [22-24] and for *fcc*-based phases in simple metals [25].

Significant arguments for the stability of tetrahedrally-packed structures are related to electrostatic (Ewald or Madelung) contributions to the crystal energy because of high packing density of atoms in Frank-Kasper polyhedra. These polyhedra are forming interpenetrating building blocks that are packed in the long-range structure with a high symmetry, as, for example, in *bcc* in the case of 1/1 approximants. The electrostatic energy of ion-ion interactions is defined usually by sums in the real and the reciprocal space providing quick convergence [26-28]. In this relation, the configuration of BZs formed by strong reflections and accommodated inside the FS is responsible for the contribution to the Ewald energy and usually prefers high symmetry polyhedra. The $oC16$ structure observed in some metals and binary alloys under pressure should be mentioned as an example [29].

## 3. Results and discussion

In this work, two groups of complex intermetallic phases are selected: with giant unit cells containing above thousand atoms and quasicrystal 1/1 approximants of the Bergman - type and Tsai - type. For our consideration we selected compounds with $s$ and $sp$ valence electron elements (non-transition elements) to allow us to calculate definitely the valence electron concentration. Structural data for compounds are given in Table 1. X-ray diffraction patterns and constructed Brillouin-Jones zones for these structures are presented in Figures 1, 2, and 3. Crystal structure descriptions are given following the Pearson notation [30].

**Table 1.** Structure parameters of several metallic compounds with giant unit cells and quasicrystal approximants. Fermi sphere radius $k_F$, ratios of $k_F$ to distances of Brillouin zone planes ½ $q_{hkl}$ and the filling degree of Brillouin zones by electron states $V_{FS}/V_{BZ}$ are calculated by the program BRIZ [21].

| Phase | $Cu_4Cd_3$ | $Mg_{28}Al_{45}$ | $CaCd_6$ | $Al_{30}Mg_{40}Zn_{30}$ | $Al_5CuLi_3$ | $Au_{15}Cd_{23}Zn_{11}$ |
|---|---|---|---|---|---|---|
| Pearson symbol | $cF1124$ | $cF1168$ | $cI168$ | $cI162$ | $cI160$ | $cI146$ |
| | | | Structural data | | | |
| Space group | $F\bar{4}3m$ | $Fd\bar{3}m$ | $Im\bar{3}$ | $Im\bar{3}$ | $Im\bar{3}$ | $Im\bar{3}$ |
| Lattice parameters (Å) | $a = 25.871$ | $a = 28.24$ | $a = 15.680$ | $a = 14.355$ | $a = 13.891$ | $a = 13.843$ |
| $V_{at.}$ (Å$^3$) | 15.67 | 19.28 | 22.95 | 18.26 | 16.75 | 18.17 |
| References | [31] | [32,33] | [34,35] | [36,37] | [38,39] | [14] |
| | | | FS – BZ data from the BRIZ program | | | |
| z (number of valence electrons per atom) | 1.43 | 2.62 | 2 | 2.3 | 2.18 | 1.69 |
| $k_F$ (Å$^{-1}$) | 1.401 | 1.590 | 1.372 | 1.551 | 1.568 | 1.403 |
| Total number BZ planes | 84 | 96 | 96 | 84 | 96 | 42 |
| hkl: $k_F/(½ q_{hkl})$ | (880):1.020 (955):1.008 (10.44):1.004 (11.33):0.979 | (14.20):1.011 (10.86) (10.10.2):1.001 | (631):1.010 (543):0.968 (701) (550) | (543):1.002 (701) (550) | (631):1.022 (543):0.988 (710) | (503):1.060 (600):1.030 (532):1.003 |
| $V_{FS}/V_{BZ}$ | 0.955 | 0.966 | 0.950 | 0.945 | 0.936 | 1.00 |

### 3.1. Giant unit cell compounds $Cu_4Cd_3$ and $Mg_2Al_3$

Complex compounds with more than 1000 atoms in the unit cell are found in $Cu_4Cd_3$ and in $Mg_2Al_3$ by Samson [31,32]. Interestingly that $Cu_4Cd_3$ compound differs considerably from isoelectronic phases in neighboring systems Cu-Zn and Au-Cd that have at this composition either *bcc* or ordered CsCl type structures [20]. The decisive factor for this difference is the atomic size ratio that does not exceed 10% for Cu/Zn or Au/Cd but is ~20% for Cu/Cd. Therefore in $Cu_4Cd_3$ instead of the simple *bcc* or CsCl structures a tetrahedrally-packed complex compound is formed.

The diffraction pattern of $Cu_4Cd_3$–$cF1124$ consists of diffraction peaks with relatively strong intensity grouped near the $2k_F$ position (Figure 1a). Below the diffraction pattern, FS-BZ constructions are shown for separate peaks and for the final BZ including 84 planes in close contact to FS (lower-right). This consideration confirms the influential role of the Hume-Rothery mechanism for the stability of a complex compound along with an atomic size factor.

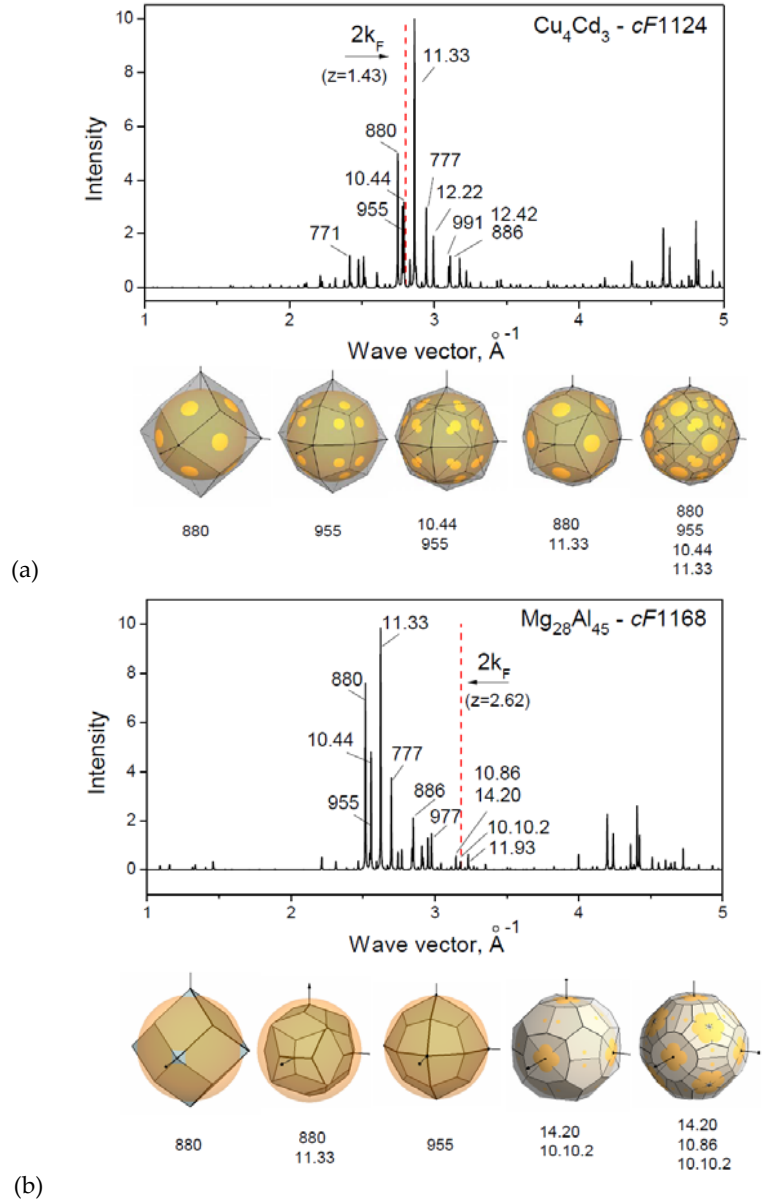

**Figure 1.** Calculated diffraction patterns for complex intermetallic phases (upper panel) and corresponding Brillouin-Jones zones with the Fermi spheres (lower panel). The position of $2k_F$ for a given valence electron number per atom z and the hkl indices of the planes used for the BZ construction are indicated on the diffraction patterns. (a) $Cu_4Cd_3$-$cF1124$; (b) $Mg_{28}Al_{45}$-$cF1168$. Structural data and FS-BZ evaluation data are given in Table 1.

A similar approach that considers Brillouin zones in the discussion of tetrahedrally-packed structures is given in [14]. It is necessary to note that for $Cu_4Cd_3$ compound same group of diffraction peaks was selected for the BZ construction in Figure 40 (ref. [14]), however the BZ form is slightly different from the construction made with the BRIZ program more accurate, as given in Figure 1a (lower-right).

The complex compound $Mg_2Al_3$ have similar structure with 1168 atoms in the unit cell as was defined by Samson [32] and re-determined recently [33]. Structural data for this phase are given in Table 1 with designation as $Mg_{28}Al_{45}$ following Pauling File data [40]. Diffraction pattern for this phase (Figure 1b) is similar to that for $Cu_4Cd_3$ (Figure 1a) with the same group of strong reflections.

*3.2. Quasicrystal approximants*

Interesting groups of tetrahedrally-packed phases represent quasicrystal approximants of 1/1-type and other types. Some of these phases have been known long before the discovery of quasicrystals, as for example $Mg_{32}(Zn,Al)_{49}$ and $CaCd_6$ [36,34] and later they were assigned as approximants of icosahedral quasicrystals (QCs). Diffraction patterns of QCs and their approximants have strong reflections at the nearly same positions and both groups of phases exist at very close alloy compositions that can be estimated as valence electron concentrations (z). It is commonly assumed that QCs and their approximants are stabilized by the Hume-Rothery mechanism [2-5] and exist in regions of z equal to ~1.8, 2 and ~2.2 - 2.3. For those z values representative QCs-approximants are considered with structural data listed in Table 1.

3.2.1. Approximants *cI*168 and *cI*162

Diffraction patterns for QCs-approximants $CaCd_6$-*cI*168 and $Al_{30}Mg_{40}Zn_{30}$-*cI*162 and FS-BZ configurations are shown in Figure 2. Both phases have similar groups of strong reflections whereas $2k_F$ positions are different and are close to the diffraction peak (631) for z = 2 and to the diffraction peaks (543), (710), (550) for z = 2.3 (Figure 2a and 2b, respectively).

Strong reflections at lower wave vector values form highly symmetrical polyhedra that are enveloped by the FS creating a contact at the edges of the polyhedron with planes (503) or vertices of the polyhedron with planes (532), (600). Due to space group symmetry $Im\bar{3}$ there is a difference in the intensity for reflections within the same (hkl) set. For example, the structure $CaCd_6$-*cI*168 results in intensity ratio ~6 for reflections (503) and (530). Assuming that only the (503) reflections contribute to the constructions of the BZ, the resulting polyhedron is of pentagonal dodecahedron type (Figure 2a, lower-left). The BZ polyhedron for (532) and (600) reflections represents a rhombic triacontahedron (Figure 2a, lower-middle). Both kinds of polyhedra are inscribed into the FS and with contacts at the vertices. Thus the polyhedra in reciprocal space are related to polyhedra in real space formed by atom clusters. It should be noted that the BZ that is in contact with the inscribed FS for $CaCd_6$ consists mainly from (631) and (361) with nearly the same intensity (Figure 2a, lower-right).

For the approximant $Al_{30}Mg_{40}Zn_{30}$–*cI*162 the next diffraction peak with indices (543), (710), and (550) forms the BZ planes in essential contact to the FS as shown in Figure 2b (lower-right). There are 84 facets in the BZ contacting the FS at the same distance. The filling of the BZ by electron states is 0.945 (see Table 1) that satisfies well the Hume-Rothery rule. Same BZ-FS configuration for $Al_{30}Mg_{40}Zn_{30}$–approximant was considered in Figure 1 (ref. [37]) with discussion of matching $2k_F$ to wave vectors of reflections (543), (710), and (550). In addition to this, it is also necessary to consider the BZs constructed with the reflections (503) and (532)/(600) that are enveloped by the FS as shown in Figure 2b (lower-left and middle).

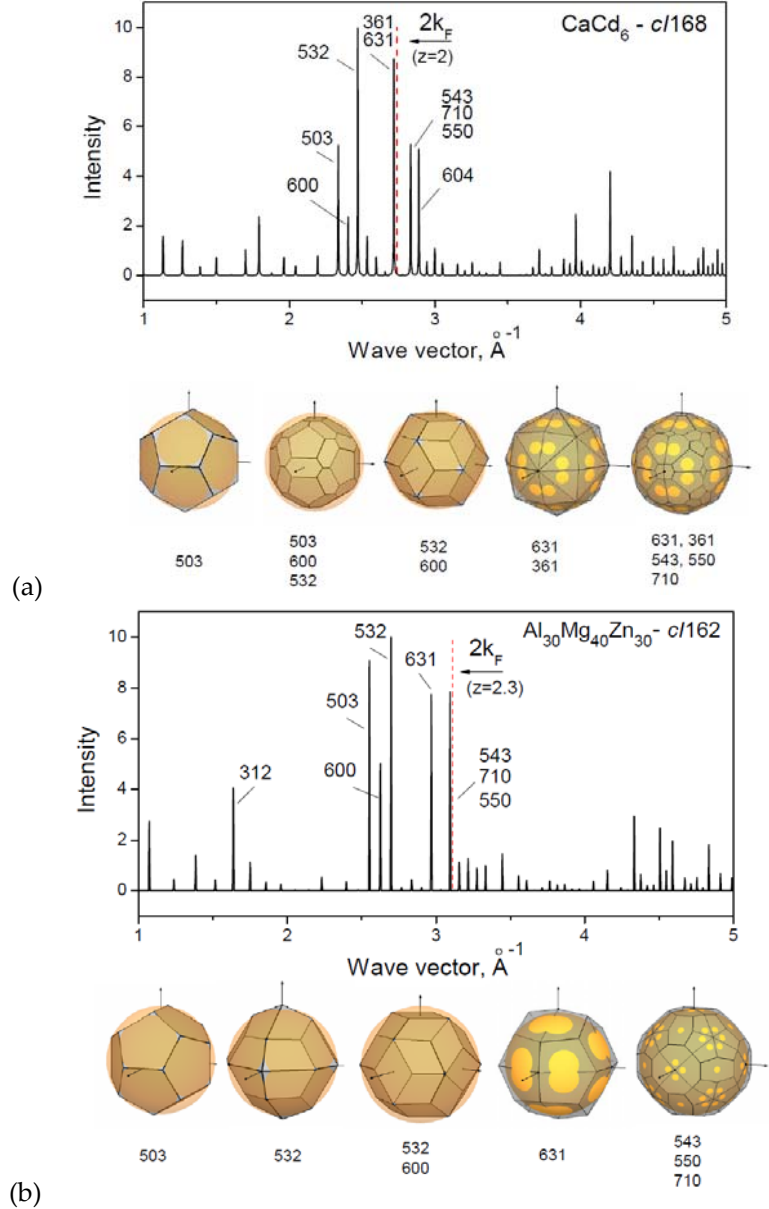

**Figure 2.** Calculated diffraction patterns for quasicrystal approximants (upper panel) and corresponding Brillouin-Jones zones with the Fermi spheres (lower panel). The position of $2k_F$ for a given valence electron number per atom $z$ and the hkl indices of the planes used for the BZ construction are indicated on the diffraction patterns. (a) $CaCd_6$-$cI$168; (b) $Al_{30}Mg_{40}Zn_{30}$-$cI$162. Structural data and FS-BZ evaluation data are given in Table 1.

3.2.2. Approximants $cI$160 and $cI$146

$Al_{30}Mg_{40}Zn_{30}$-$cI$162 and $Al_5CuLi_3$-$cI$160 have much similarity in their crystal structures, both belonging to the Bergman-type approximants. Diffraction patterns for both phases are similar, as can be seen from Figures 2b and 3a. The only difference is in the $2k_F$ position: for $Al_5CuLi_3$ it is close to the reflection (631) while for $Al_{30}Mg_{40}Zn_{30}$ it is close to reflections (543), (710), and (550), as shown in Figures 2b and 3a (lower-right), respectively. It should be noted, that because of space group $Im\bar{3}$ the intensity of the (613) reflection is ~100 times less than that of (631) and the only (631) set is participating in the BZ construction. For $Al_5CuLi_3$ the BZ filling by electron state is 0.936 matching the Hume-Rothery criteria. Besides this FS-BZ planes contact it is also necessary to consider the FS overlap with the inner zones such as (503)-dodecahedron and rhombic triacontahedron formed by

(532) and (600) planes that should give a significant contribution to the reduction of the electronic energy.

Our next example of approximants – the phase $Au_{15}Cd_{23}Zn_{11}$-$cI$146 was found recently [14] with the valence electron number $z = 1.694$. For this z value, $2k_F$ position is close to (532) and slightly overlaps the (503) and (600) reflections as shown in Figure 3b. The resulting BZ (Figure 3b, lower-right) is completely filled by electron states. It should be noted that BZ construction for this structure was reported in Figure 34 (ref. [14]) however BRIZ program gives more accurate construction. Interestingly, for this phase with $z\sim1.7$ there are no other diffraction peaks with comparably high intensity in contrast to the phases with higher z values.

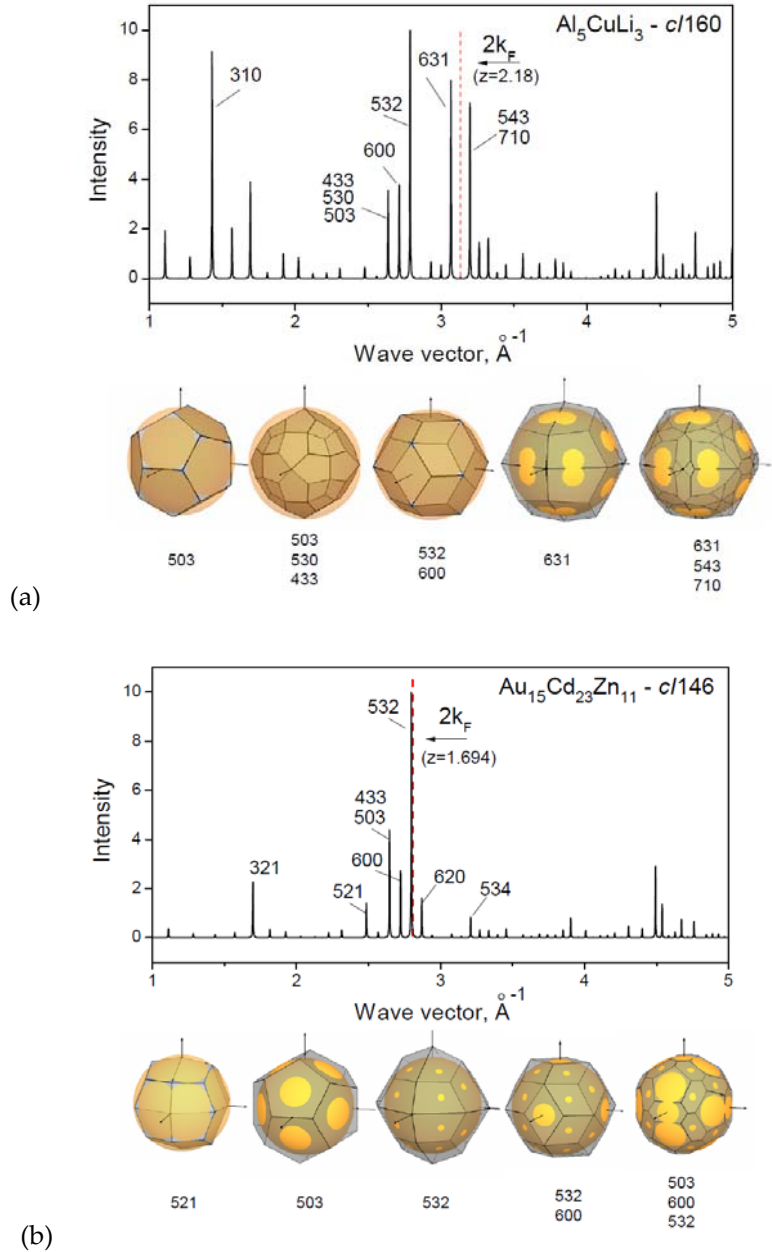

(a)

(b)

**Figure 3.** Calculated diffraction patterns for quasicrystal approximants (upper panel) and corresponding Brillouin-Jones zones with the Fermi spheres (lower panel). The position of $2k_F$ for a given valence electron number per atom $z$ and the hkl indices of the planes used for the BZ construction are indicated on the diffraction patterns. (a) $Al_5CuLi_3$-$cI$160; (b) $Au_{15}Cd_{23}Zn_{11}$-$cI$146. Structural data and FS-BZ evaluation data are given in Table 1.

*3.3. Similar structural features of intermetallics with giant unit cells and approximants*

The complex intermetallic phases discussed in the present paper have similar local atomic arrangement based on tetrahedrally-packed Frank-Kasper polyhedra. Relations of atomic configurations organized in different structural types lead to relations in a form of Brillouin-Jones zones. Examples of such similarity provide the BZ constructions for $Mg_{28}Al_{45}$–$cI1168$ and $Al_{30}Mg_{40}Zn_{30}$–$cI162$ shown in Figure 1b and 2b (lower-right). Diffraction plain sets for BZ construction contacting the FS are following:

$Mg_{28}Al_{45}$    (10. 8 6)    (14. 2 0)    (10. 10. 2)
$Al_{30}Mg_{40}Zn_{30}$    (5 4 3)    (7 1 0)    (5 5 0).

From the comparison of these sets it can be concluded that the former cell is related to the latter cell as 2×2×2 by lattice parameter. In this case the number of atoms would be 162 × 8 = 1296 that can be reduced to 1168 by vacancies. It should be noted that for $Mg_{28}Al_{45}$ a reflection (10.10.0) expected from this transformation also exists, but has very weak intensity, and for the construction of the BZ the next reflection (10.10.2) was taken. Formation of a complex supercell by multiplication of lattice parameters was demonstrated by a classical example of γ-brass phase $Cu_5Zn_8$. The structure $cI52$ is formed from *bcc* with the 2×2×2 increase of lattice parameter and an induction of vacancies with some atomic movement that results in an appearance of additional peaks close to the FS, as was discussed by Jones [12]. Structural presentations of complex intermetallics as multi-fold supercells of basic metallic close-packed lattices were considered in [25,41].

Structural relations of alloy phases that are built with similar short-range clusters are visible by comparison of BZs related to strong diffraction peaks. For $Al_5CuLi_3$-$cI160$ the outer BZ with faces (503), (600) and (532) shown in Figure 3b (lower-right) represents a polyhedron that looks like Brillouin-Jones zone of quasicrystals i-AlMgCu and i-AlLiCu constructed from first strong peaks [6]. It should be noted, that BZs with this type of polyhedron with faces (503), (600) and (532) exist for all approximants discussed where the only difference is in the location either outside or inside the FS. In both cases the FS-BZ interaction is significant for the crystal structure energy.

**4. Conclusion**

Structurally complex alloy phases with tetrahedrally-packed polyhedra of Frank-Kasper type are discussed with reference to the free-electron model of Brillouin zone – Fermi sphere interactions. Electron energy reduction originates through the contact of the FS to the BZ planes as assumed usually for classical Hume-Rothery phases with relatively simple structures. On their diffraction patterns, complex structures display several peaks of strong intensity that produce Brillouin zone polyhedra enveloped by the FS with the contact to some edges or vertices. These FS-BZ configurations should be taken into account for the estimation of electron energy as a convincing reason for stability of such complex intermetallic structures. Complex tetrahedrally-packed intermetallic phases consist of successive atomic building blocks arranged as symmetrical polyhedra. The Brillouin-zone constructions for these structures reveal successive BZ polyhedra with nearly same form as atomic clusters in the real space. These effects should be regarded by the theoretical considerations.

**Acknowledgments:** The authors gratefully acknowledge Dr. Olga Degtyareva for valuable discussion and comments. This work is supported by the Program "The Matter under High Pressure" of the Russian Academy of Sciences.